\documentclass[review,3p, times]{elsarticle}

\usepackage{lineno,hyperref}

\usepackage[utf8]{inputenc}    
\usepackage[T1]{fontenc} 
\usepackage[english]{babel} 
\usepackage{graphicx}
\usepackage{enumitem} 
\usepackage{hyperref} 
\usepackage{upgreek}
\usepackage{siunitx}
\usepackage{xcolor}
\usepackage{float}
\usepackage{setspace}
\usepackage{amsmath,amsfonts,amssymb}

\graphicspath{{images/}}

\DeclareUnicodeCharacter{2212}{-}

\journal{Radiation Physics and Chemistry}

\begin{document}

\begin{frontmatter}
	\title{Optimization of GEM detectors for applications in X-ray fluorescence imaging.} 
	\author{G. G. A. de Souza$^{1}$\corref{corauthor}}
	\cortext[corauthor]{Corresponding author}
	\ead{geovane.souza@usp.br}
	\author{H. N. da Luz$^{2}$, M. Bregant$^{1}$}
	\address{1-Instituto de F\'isica, Universidade de S\~ao Paulo\\Rua do Matão 1371, 05508-090 Cidade Universitária, São Paulo, Brazil \\
		2 - Institute of Experimental and Applied Physics, Czech Technical University in Prague \\ Husova 5, 110 00 Prague 1, Czech Republic
	}
	
	\begin{abstract}
		In this work a set of simulations that aim at the optimization of Micropattern Gaseous Detectors (MPGD) for applications in X-ray fluorescence imaging in the energy range of 3 -- \SI{30}{\kilo\electronvolt} is presented. 
		By studying the statistical distribution of electrons from interactions of X-rays with gases, the energy resolution limits after charge multiplication for \SI{6}{\kilo\electronvolt} X-ray photons in Ar/CO$_2$(70/30) and Kr/CO$_2$(90/10) were calculated, resulting in energy resolutions of 15.4(4)\% and 14.6(2)\% respectively. These two mixtures were  studied in simulations to evaluate the advantages of using krypton-based mixtures to reduce the presence of escape peaks in fluorescence spectra. A model to evaluate the X-ray fluorescence from the conductive materials inside the detectors was implemented, serving as a tool to estimate the extent of contamination of fluorescence spectra when using copper or aluminium layers in the material composition of MPGDs.
	\end{abstract}

\end{frontmatter}  

\section{Introduction}


Micro-pattern Gaseous Detectors~(MPGD)~\cite{Car12, Alt02, ALICEUP, CMS15} have earned their place in high energy and particle physics experiments due to their performance as charge multiplication devices in tracking and particle identification detectors. MPGDs are a key component in the construction and performance of this type of detectors, but have also spread their use to other applications, showing promising results in the detection of X-rays while providing a good position sensitivity, necessary, for example, for cultural heritage studies\cite{Bar21, Zie13, Ana15} that can even be complemented with tomography techniques \cite{Car19}. The main advantage of using such technology is the larger area coverage ($\geq \SI{100}{\centi\meter\squared}$) when compared to solid state detectors, that spares the need to thoroughly scan the samples. These techniques have undergone a steady development in the past few years, but there is room to optimize parameters of the detector to obtain more reliable results in the energy range of 3 -- \SI{30}{\kilo\electronvolt}, which covers the K or L emission lines of most elements.

The Gas Electron Multiplier (GEM) \cite{Sau16} is one of the most popular MPGD. It consists of a \SI{50}{\micro\m} thin kapton foil, copper clad on both sides and perforated with small holes with a diameter of \SI{50}{\micro\meter} laid out in a triangular array with a pitch of \SI{140}{\micro\meter}. When immersed in a gas, a suitable voltage difference between the two conductive layers generates inside the holes an electric field that is well above the Townsend limit for electron multiplication in the gas. The free electrons in the gas, generated by the incoming ionizing radiation, are focused towards the holes where they are multiplied and transferred to the opposite side towards the next GEM foil or the readout anode. A more detailed description is provided further below and more details on these detectors can be found in dedicated literature \cite{Sau16,Sau14,gddlab}.

When using GEM detectors to measure the energy and position of the interaction of X-rays at such low energy ranges, there are well documented effects \cite{Min17, Min20, Geo19} that cause limitations in the detection efficiency of the detectors, in the position resolution and in the capability of quantifying (or even distinguishing) different elements from a sample. In fact, the presence of fluorescence peaks caused by the materials present in the detector, such as the copper used in the conductive layers of the GEM foils; the escape peaks caused by the loss of part of the initial energy in the form of fluorescence X-rays emitted by the absorbing gas and escaping the detector, among other effects have been confirmed in many experimental works. The aim of this work is to discuss possibilities of coping with --- or mitigating --- these effects. This work studies the standard materials and gases typically used in GEM-based detectors, which are copper clad foils and argon-based mixtures and compares their performance with results that have been tested and published by other groups working in this topic. The main objective is to provide a deeper understanding of the physical processes and the development of simulation tools to evaluate new solutions. All the analytical and simulation results presented are derived from known Physics applied to the detector setups described.

A set of simulations is presented as proposals for the optimization of GEM-based detectors for X-ray fluorescence~(XRF) and imaging. The simulations were done using the Garfield++ libraries\cite{Garfieldplus} and standard C++ code. In section~\ref{sec:quantumeff}, results regarding the photon conversion efficiency and the energy resolution limits that can be achieved with GEM-based gaseous detectors in the desired energy range are presented. Section~\ref{section:escape} shows results related to the photon escape probability and proposes possibilities to reduce the escape peak contribution to the energy spectrum. Since this work is focused in X-ray fluorescence analysis and imaging, a model that describes the extra fluorescence peak generated by the presence of copper in the GEM foils was also studied, as described in section~\ref{sec:coppercontamination}. 

\section{Detection efficiency}
\label{sec:quantumeff}
A typical X-ray fluorescence imaging setup consists of a triple-GEM detector that is operated at atmospheric pressure, using neon, argon or krypton mixed with carbon dioxide at different relative concentrations. Other gas mixtures can be used, but in this work we are focused on the mixtures that are more common in works on this topic. 

 This GEM-based detector consists of a sensitive volume where the radiation is absorbed. This region is under a mild uniform electric field that prevents the electron-ion pairs from recombining, making the electrons drift towards the stack of GEM foils. The uniform electric field is often referred as the drift field and the sensitive or absorption region is also often referred as drift region. Thanks to the geometry of the GEM foils, the electric field lines are focused  towards their holes, where the field is high enough to generate a Townsend avalanche that can be further transferred to the next GEM foil. In the end, the induction (or extraction) field removes the electrons from the holes and brings them to the readout electrode. This induction field is below the multiplication limit, but must be strong enough to prevent the electrons from being collected by the bottom electrode of the last foil.

\begin{figure}
	\centering
	\includegraphics[width=6cm]{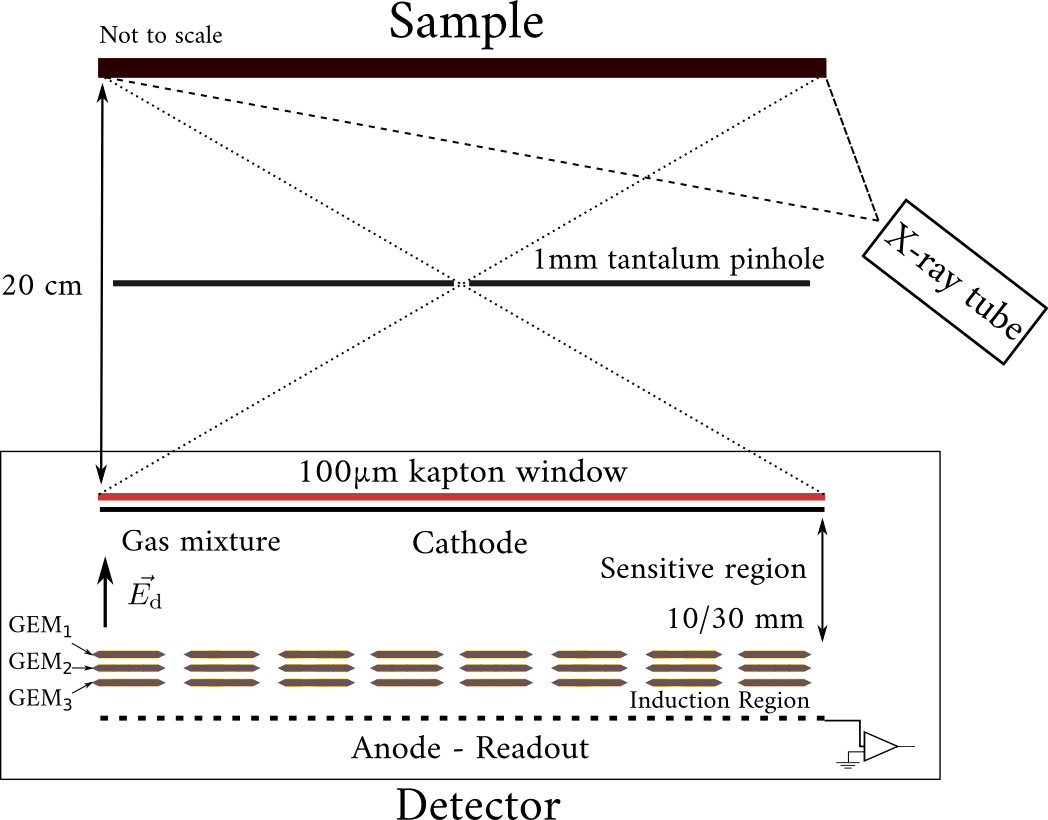}
	
	\caption{Detector setup used for X-ray fluorescence. Not to scale.}
	\label{detec_setup}
\end{figure}

The detector is operated with a charge gain close to 2$\times10^4$. The sample is irradiated by an X-ray source and, by using a pin-hole geometry, an image of the sample is projected on the detector as shown in figure \ref{detec_setup}. After the charge multiplication inside the detector, the electrons are collected by a segmented strip readout (typically 256 strips for each dimension) and the position and energy of the event are reconstructed using signal processing tools (more details can be found in \cite{Geo19}). There are \SI{20}{\centi\meter} of air and a \SI{100}{\micro\meter} kapton window between the source and the detector, for which a simple calculation (using the Beer-Lambert law or the on-line tables \cite{XCOM,Hen93}, which use the same law) shows that most of the radiation below \SI{3}{\kilo\electronvolt} is absorbed, leading to a limitation on the lowest energy that can be detected. Two different configurations of the detector's sensitive region,  where the conversion of the photons in electron-ion pairs takes place, were considered: one with \SI{1}{\centi\meter} and the other with \SI{3}{\centi\meter}.  

\begin{figure}
	\centering
	\includegraphics[width=12cm]{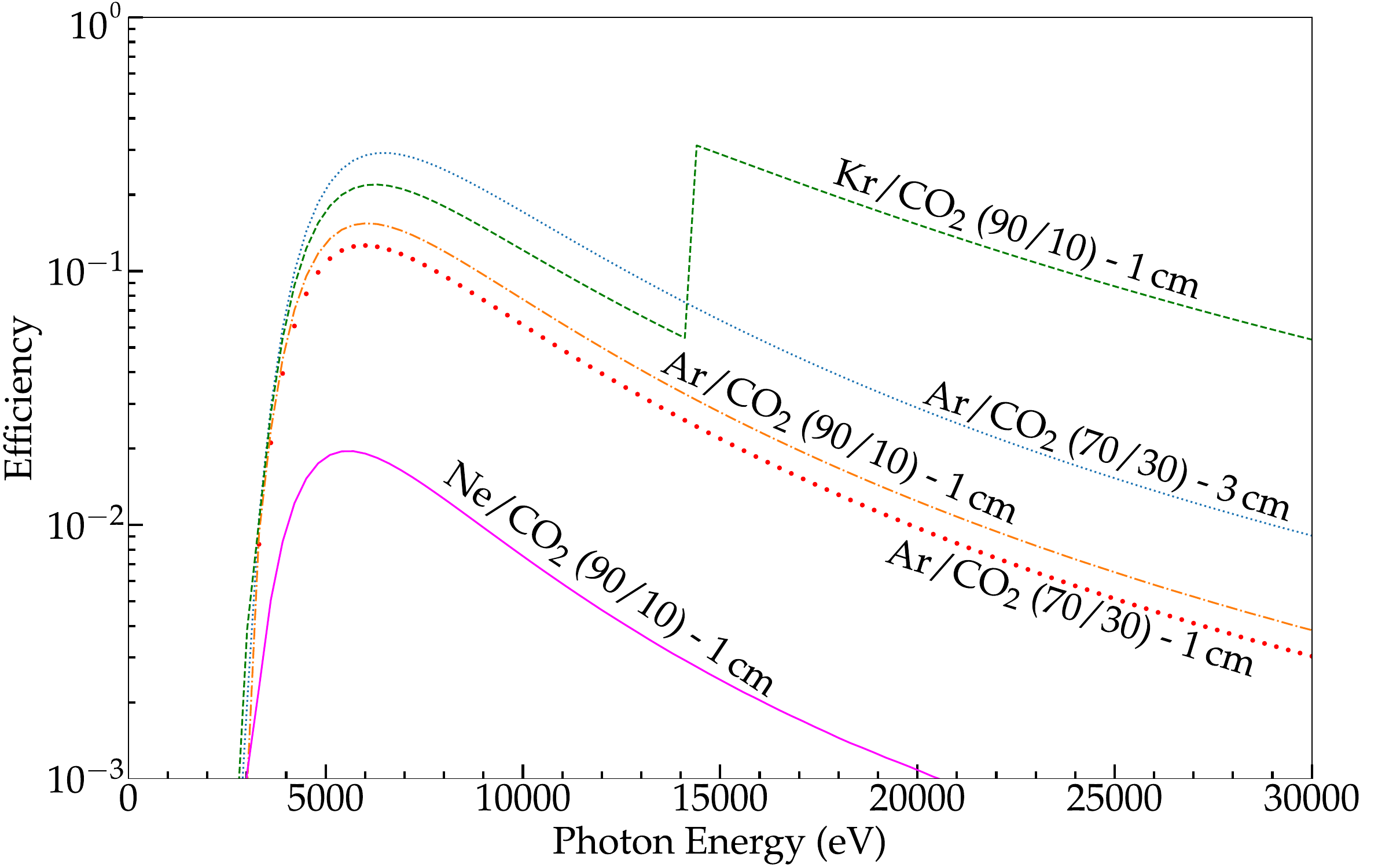}
	
	\caption{Computed detection efficiency for three different noble gaseous mixtures.}
	\label{gas_eff}
\end{figure}

Figure \ref{gas_eff} shows the simulated detector efficiency (calculated using the data tables from \cite{Hen93}) for three different gas mixtures at a 90/10 proportion and 1~cm deep sensitive region (Ar/CO$_2$, Ne/CO$_2$ and Kr/CO$_2$) and two curves for Ar/CO$_2$(70/10) for  1- and 3~cm deep sensitive region. The attenuation due to the air between the sample and the detector and due to the kapton window are included in the calculations. Comparing between the different configurations, argon and krypton show an efficiency of the same order of magnitude up to \SI{14}{\kilo\electronvolt}. Above this energy, krypton shows a higher efficiency due to the K absorption edge. Neon has a detection efficiency that is approximately one order of magnitude smaller than the other gases as expected from its lower atomic number.

\begin{figure}[h]
	\centering
	\includegraphics[width=8cm]{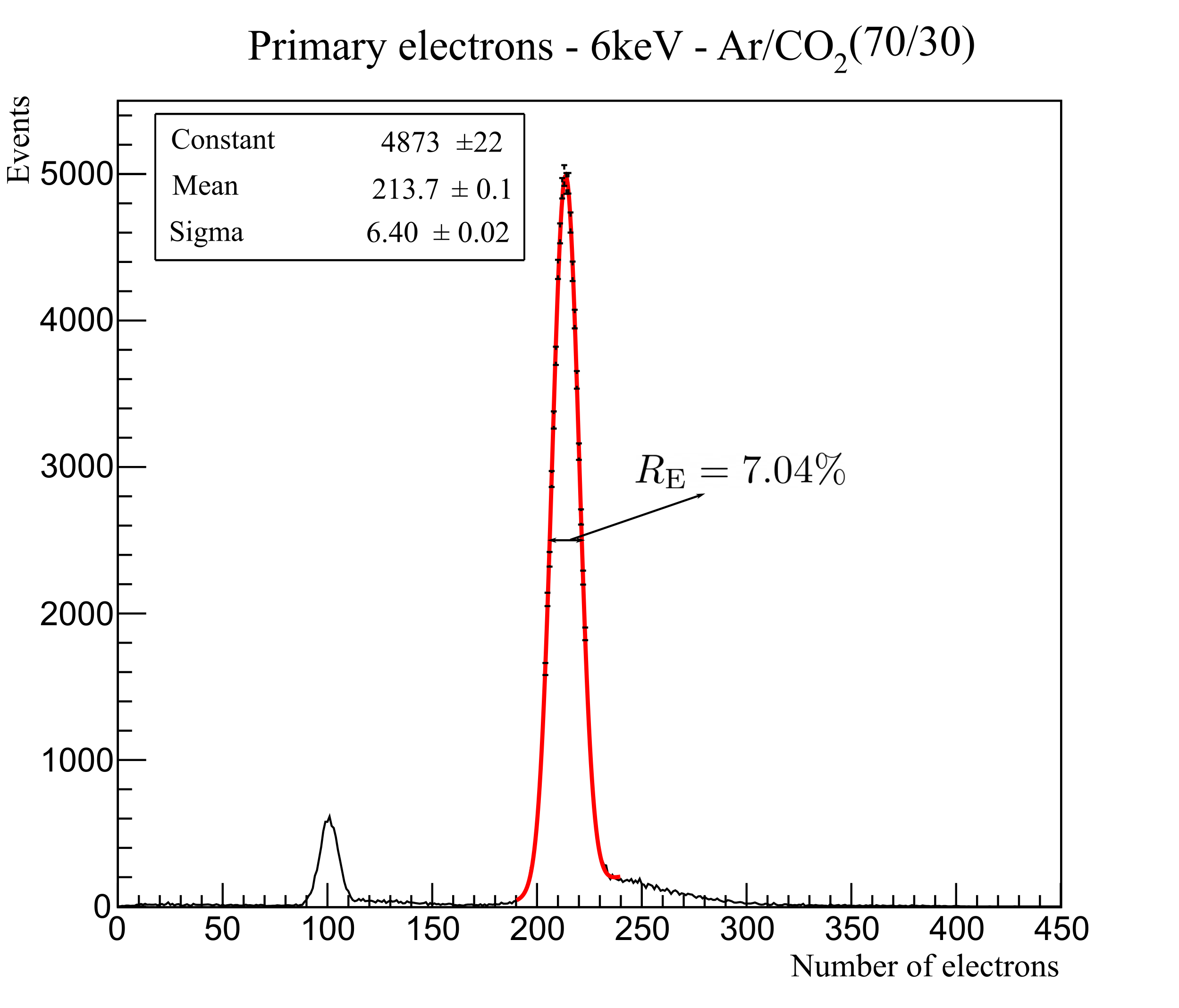}
	\includegraphics[width=8cm]{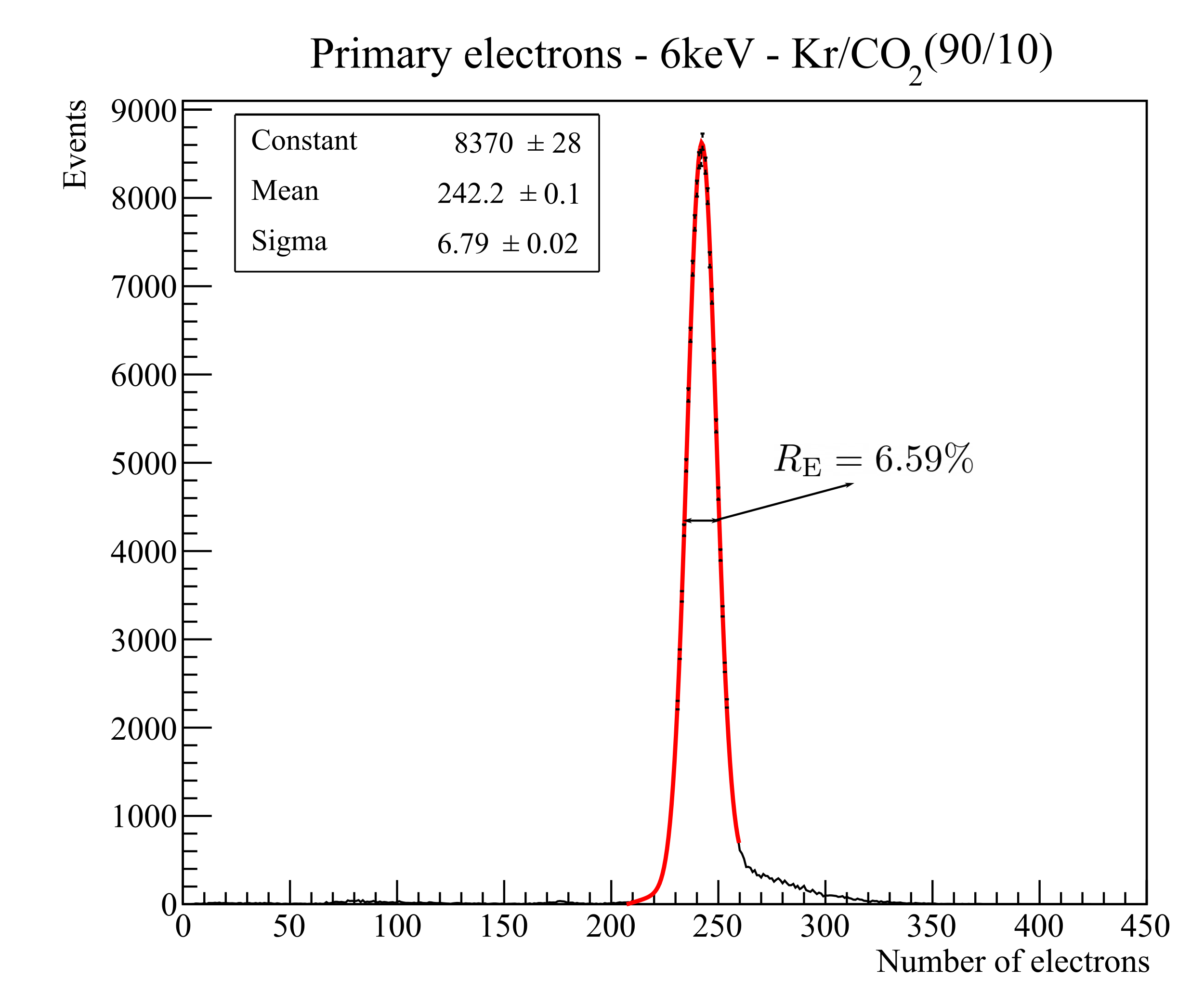}
	\caption{Number of electrons generated in the sensitive region by monoenergetic photons with \SI{6}{\kilo\electronvolt} for argon mixture~(left) and krypton mixture~(right) and a sensitive region with a depth of 1\,cm. The total number of photons interacting was $10^5.$ The intrinsic energy resolution of the gas mixtures is given by the full width at half maximum (FWHM).}
	\label{primary_spectrum}
\end{figure}

To study the statistical fluctuations in the electron-ion pair creation inside the detector and how they influence the energy resolution, the HEED \cite{Smi05} package, which is available inside the Garfield++ program, was used to simulate the interaction of monoenergetic photons inside a gaseous volume filled with Ar/CO$_2$ or Kr/CO$_2$ mixtures. The simulation was done for $10^5$ photons absorbed by the detector.  Figure~\ref{primary_spectrum} shows the distribution of the number of primary electrons~($Q_{\mathrm{p}}$) generated by the \SI{6}{\kilo\electronvolt} photon interaction with Ar/CO$_2$(70/30) and Kr/CO$_2$(90/10). Both sensitive regions are simulated at atmospheric pressure, with dimensions of 10 cm width $\times$ 10 cm depth $\times$ 1 cm height. The number of primary electrons is distributed around a mean value that is proportional to the deposited energy. The values obtained are 214 and 242 primary electrons for the argon and the krypton mixtures, respectively. These values are in agreement with those expected by dividing the incident X-ray energy by the average energy needed for an electron-ion pair creation in argon, krypton and carbon dioxyde published in the literature (ex.:~\cite{Ber12,Bor91,Jar83,Sip76}). The secondary peak, at lower energy in the plot for the argon-based mixture, is the argon escape peak. The width of the main peaks reflects the intrinsic resolution of the gas mixture, not the final resolution of the detector, which is influenced by other factors that will be discussed in the following lines. 

The number of primary electrons generated after the first ionization is described by Poisson processes. However, the distribution is narrower than what is predicted by purely statistical considerations due to the fact that the processes generating the primary cloud are not completely independent from each other. This difference is corrected by the Fano factor \cite{Leo}. If the number of primary electrons followed a pure Poisson distribution, the standard deviation of the number of primary electrons ($\sigma$), expected for \SI{6}{\kilo\electronvolt} interacting in argon and krypton would be the square root of the number of primary electrons, i.e., 15.2 and 16.6, respectively. However, in these simulations a $\sigma$ of 6.4 and 6.79 was obtained, corresponding to a Fano factor of 0.209$\pm$0.005 and 0.217$\pm$0.005, in agreement with other publications~\cite{Kas84,Rib83}.

\begin{figure}[h]
	\centering
	\includegraphics[width=7.5cm]{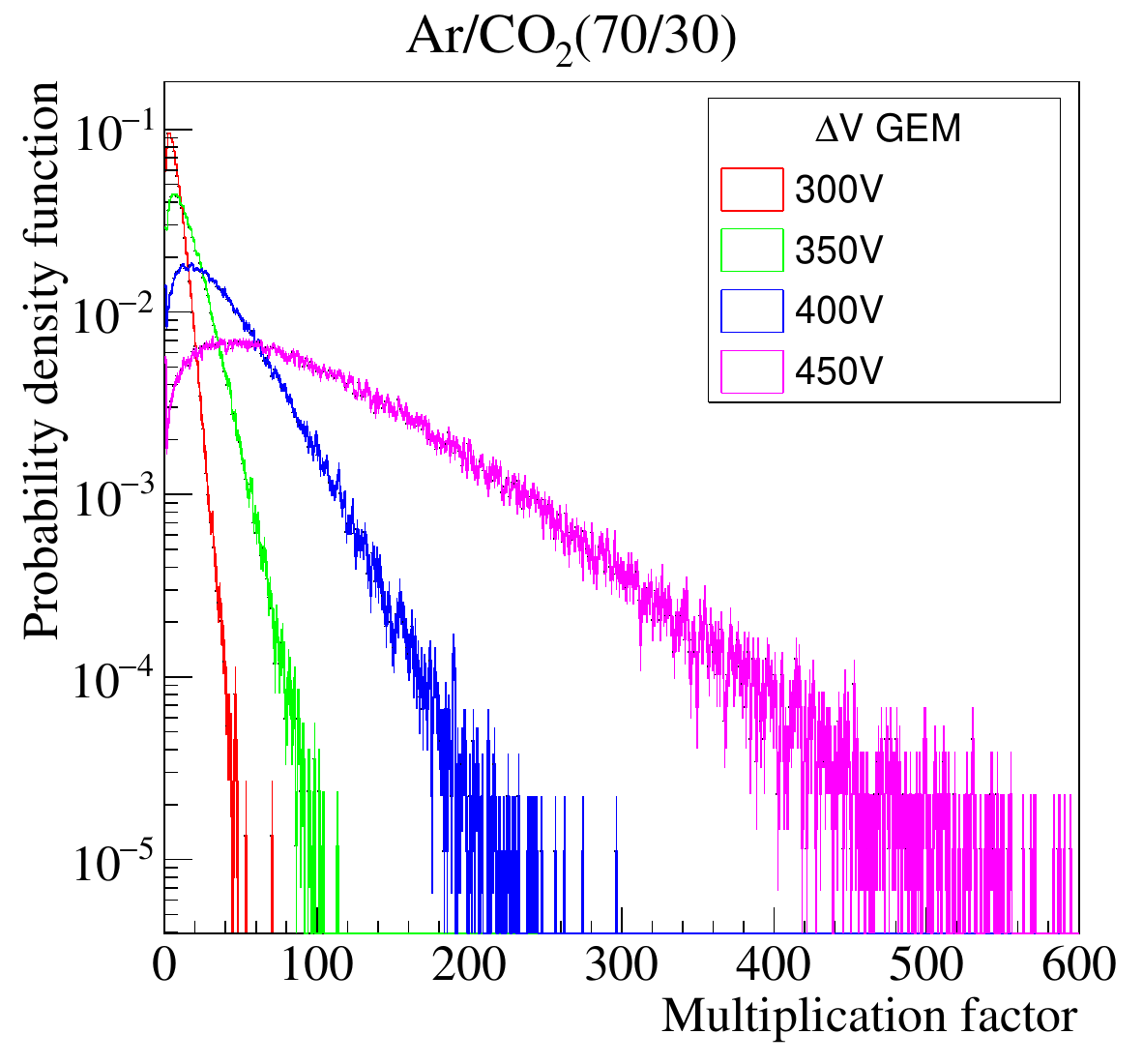}
	\includegraphics[width=7.5cm]{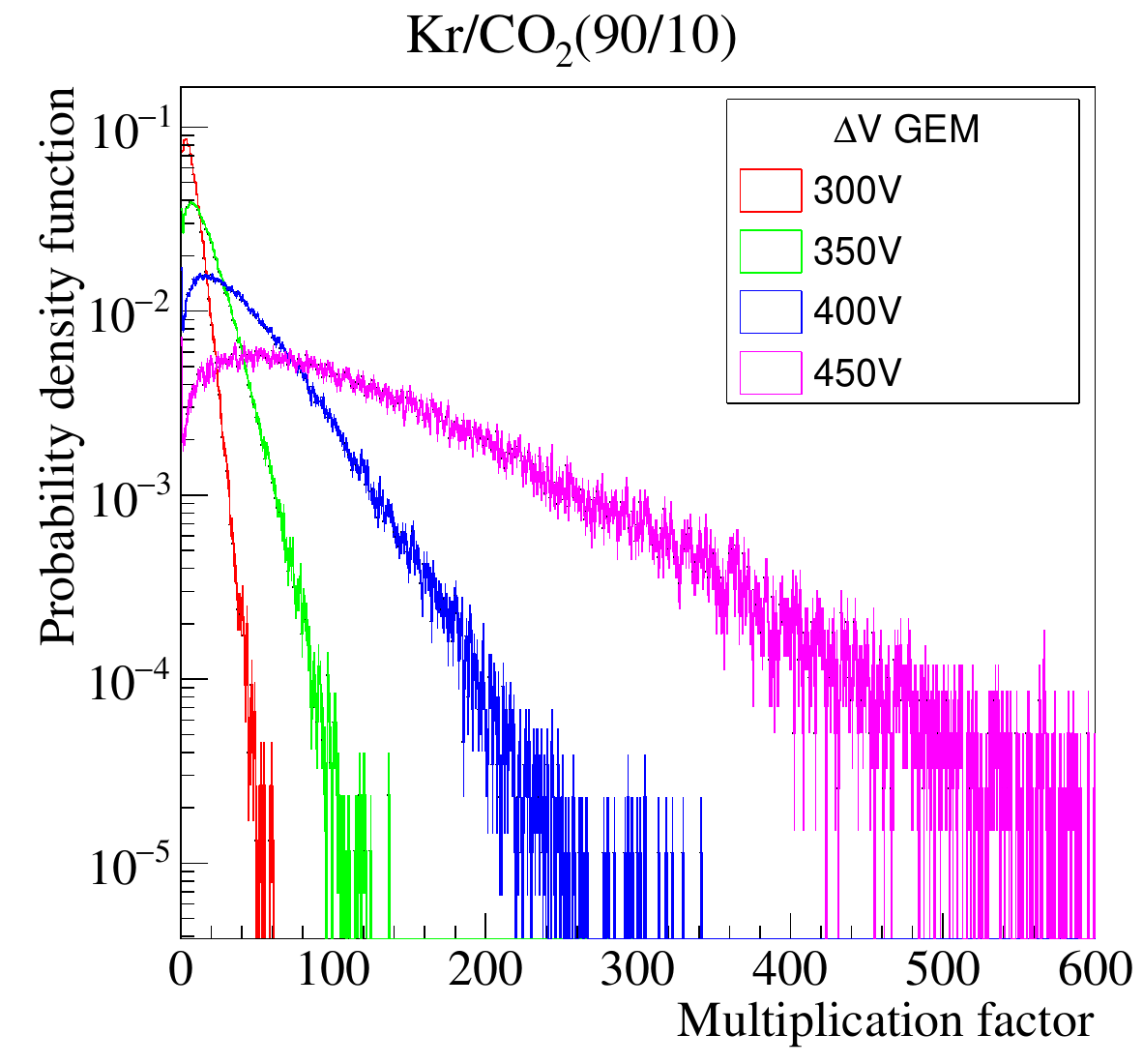}
	\caption{Simulation of the number of electrons produced after the GEM multiplication of a single primary electron.}
	\label{arco270_30}
\end{figure}

In any case, the more primary electrons are generated, the smaller are the relative fluctuations in this number, resulting in a better resolution for higher energies. The multiplication mechanism to amplify the primary charge (Townsend avalanche) contributes to increase the statistical fluctuations and will generate a final energy distribution which is wider when compared to the distribution of the primary electrons. Figure~\ref{arco270_30} shows a simulation of the multiplication factor distribution for a single electron obtained with a single GEM foil at different operating voltages. The multiplication factor is defined as the amount of charge collected by the readout electronics of the detector ($Q_{\mathrm{a}}$) divided by the primary charge ($Q_{\mathrm{p}}$). It corresponds to the effective gain of the detector if applied to the primary electron clouds. However, in this case we are simulating the response of the detector to one single electron. The simulated drift field ($E_d$) was set to \SI[per-mode=symbol]{400}{\volt\per\centi\meter} while the induction field ($E_i$) was kept at \SI[per-mode=symbol]{4000}{\volt\per\centi\meter}. These are typical magnitudes of the fields used in this type of detector and are related with the standard operational conditions of the detector, mainly in terms of maximization of the collection and extraction efficiencies of the electrons entering and leaving the holes.

\begin{figure}[h]
	\centering
	\includegraphics[width=8cm]{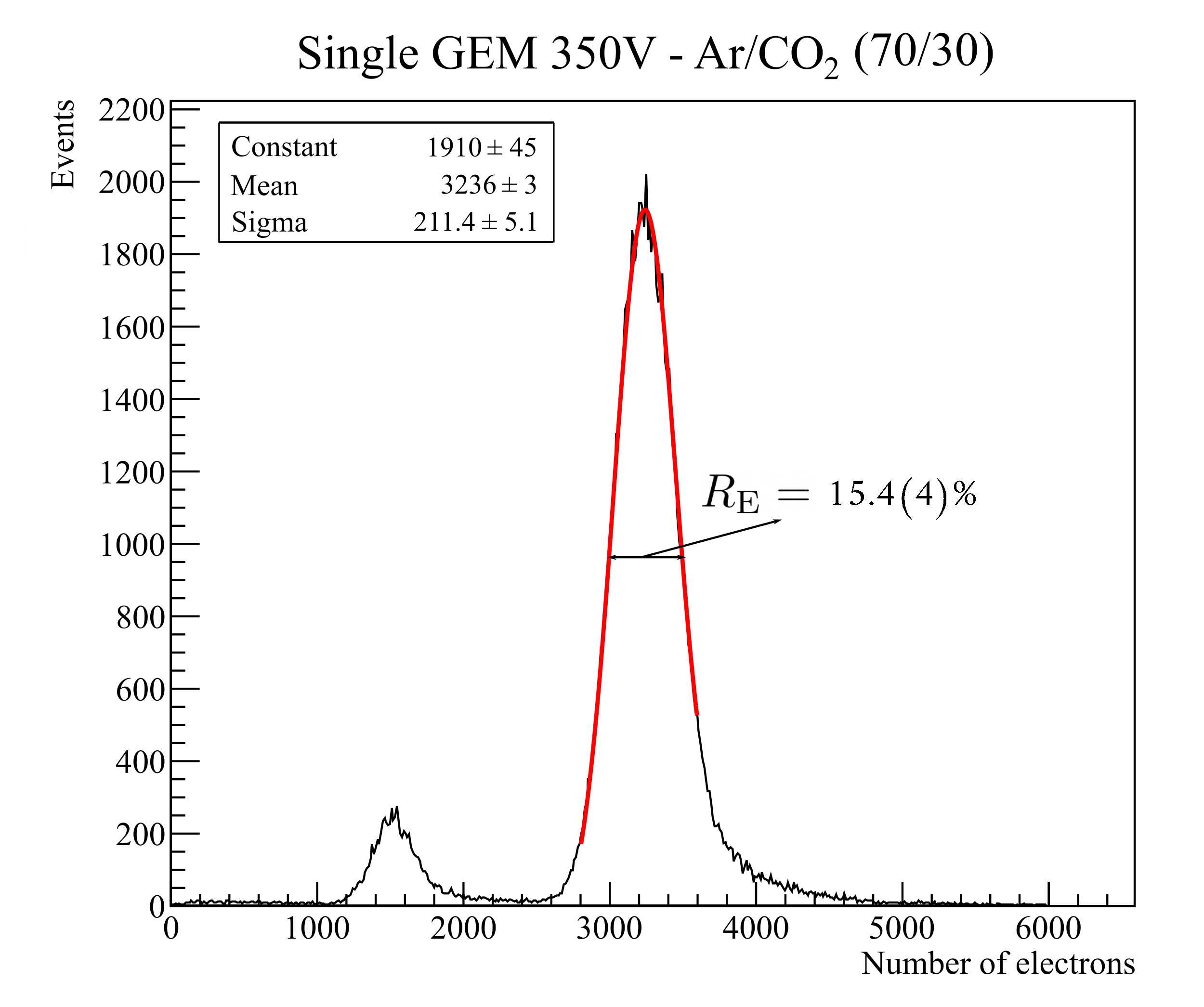}
	\includegraphics[width=8cm]{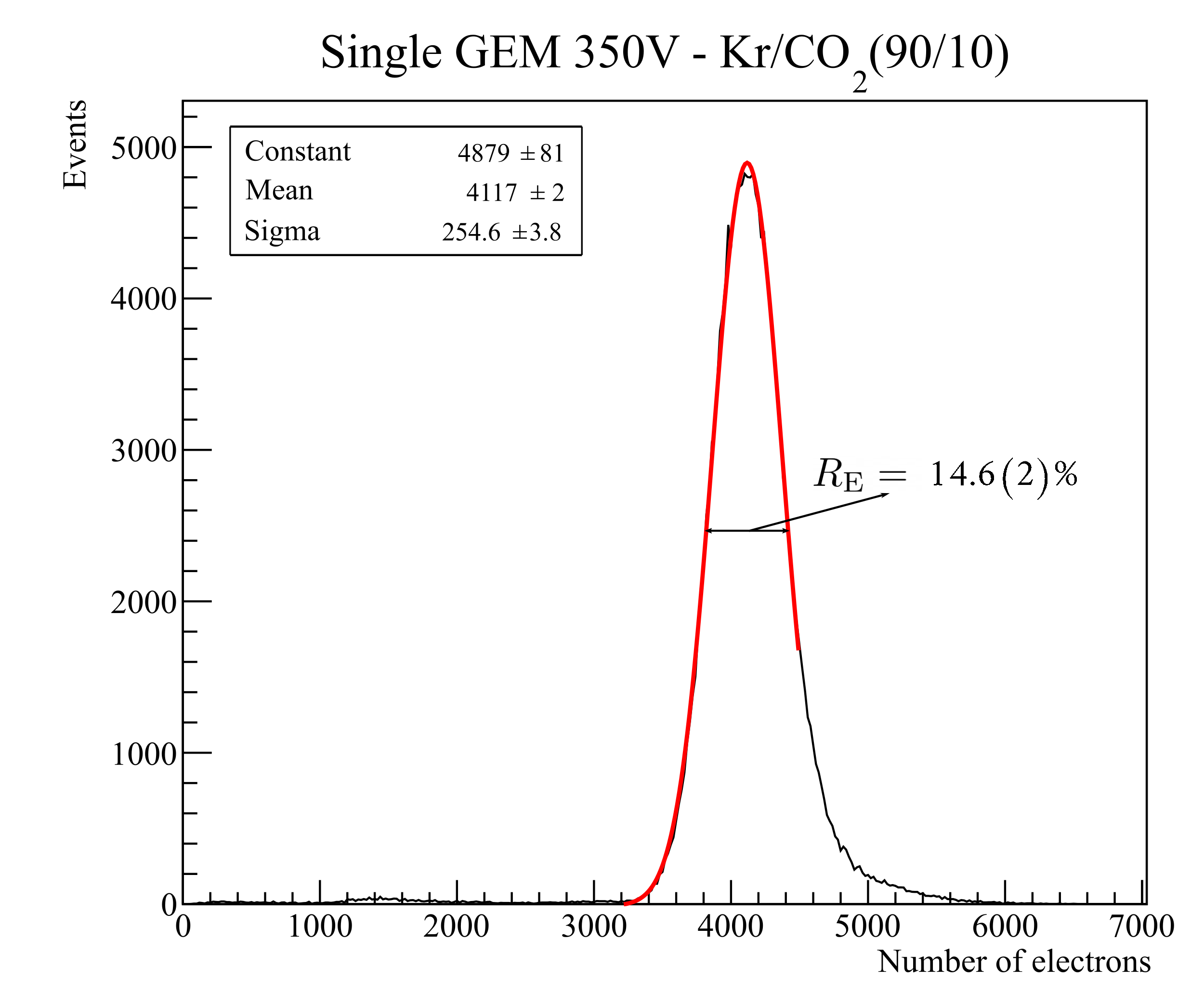}
	\caption{Number of electrons (generated by \SI{6}{\kilo\electronvolt} photons) collected at the anode after multiplication by a single GEM foil. The resolution is the FWHM.}
	\label{single_stack}
\end{figure}

The Y-axis in figure \ref{arco270_30} shows the probability density of each multiplication factor at different voltages across the holes of the GEM foils. These probability density distributions allow to define probabilistically a multiplication factor to each primary electron used to make the plots of figure \ref{primary_spectrum}, producing a histogram of the final distribution of the number of electrons per X-ray, collected by the readout electrode of the detector. Figure~\ref{single_stack} shows the electron distribution after multiplication, using the curves for \SI{350}{\volt}, for the argon and krypton mixtures. As expected, after multiplication, the energy resolution is degraded due to the statistical process that takes place in the avalanche process. The same procedure was made for a stack of two and three GEM foils, however, no considerable difference in the energy resolution was found in the simulation, reflecting the smaller statistical fluctuations due to the much larger number of electrons now available. The value obtained shows the best energy resolution that can be achieved with ideal GEM foils for those specific gas mixtures and X-ray energy, neglecting other effects such as electronic noise or detector non uniformity. When working in the laboratory, additional GEM foils will influence the energy resolution obtained since they will change the detector behavior such as the discharge probability or the ion backflow, but mainly will increase the signal-to-noise ratio. 

For \SI{6}{\kilo\electronvolt} the value for the energy resolution before charge multiplication in Ar/CO$_{2}$ was close to 7\% (FWHM), at atmospheric pressure. After charge multiplication, the value for energy resolution degrades to 15\% (FWHM). This may cause uncertainties in the identification of elements with adjacent atomic numbers when using GEM detectors for XRF applications. The same effect in Kr/CO$_2$(90/10) was simulated and since the Fano's factors and effective gain achieved are similar for Krypton and Argon \cite{Kas84,Rib83,Ort03} are similar, the resolution degradation was similar.
Reference~\cite{Geo19} contains an example of a typical pulse heigh distribution obtained with a Triple-GEM detector irradiated with a $^{55}\mathrm{Fe}$ radioactive source. This distribution shows that other effects related to the non-ideal nature of the detector are influencing the energy resolution, as stated before, but the simulation shows the same features obtained experimentally.

\section{Photon escape probability}
\label{section:escape}

In the energy spectra of the argon-based mixture shown in figures \ref{primary_spectrum} and \ref{single_stack} the escape peak can be clearly seen, with lower intensity and less energy than the main peak. It is generated by events where the energy of the incident photons is not completely deposited inside the detector. When X-rays interact with the gas in the sensitive region of the detector, electrons from the inner shells are ejected mainly by photoelectric effect. These photoelectrons will deposit the excess of energy in the detector through ionization. The energy spent to remove the electron is deposited in the detector when the excited ion returns to its fundamental state, it can release a fluorescence X-ray photon that can interact again in the sensitive volume. However, if this X-ray escapes the detector, the collected energy will be the energy of the incident X-ray subtracted by the energy of the escaping X-ray. This is a very well known effect in every detector, that results in secondary peaks in the energy spectrum at an energy equal to the incident X-ray minus the difference of energies between the two shells giving origin to the electron transition. 
 
For argon, the energy difference between the L and the K shells  is \SI{3}{\kilo\electronvolt} and for krypton, between the M and L shells, it is  \SI{1.59}{\kilo\electronvolt} (considering incident energies below \SI{14}{\kilo\electronvolt}). The escape probability for krypton is much smaller due to its lower energy and lower fluorescence yield for the L-shell\cite {Coh87}. For each characteristic X-ray line in the detected energy spectrum an escape peak is generated with a well known energy defined by the filling gas. This poses a challenge when trying to identify elements by spectral analysis, due to the overlapping of the escape peaks with the fluorescence lines emitted by the sample under study~\cite{Geo19}. To better understand the extent of such effects, one can simulate the probability of a photon escaping as a function of the height of the sensitive region and the type of gas used inside the detector.

From the Beer-Lambert equation, which describes the attenuation of the intensity of photons in matter, and using the atom fluorescence yield $\omega$, which provides the fraction of vacancies of atomic shells that are filled through a radiative process, the probability of a fluorescence photon to be produced and to leave the detector without depositing its energy can be determined by:

\begin{equation}
P(x) =\omega e^{-\frac{x}{\lambda}} ,
\end{equation}

\noindent where $x$ is the length traveled by the photon inside the medium and $\lambda$ is the attenuation length for a given energy in the same medium. The attenuation length is determined from the X-ray interaction cross sections for the different possible effects of X-rays in matter. In this work we performed a Monte-Carlo calculation using the XCOM database\cite{XCOM}. The two different detector configurations mentioned above were considered: the first one with a \SI{1}{\cm} absorption region and the second with \SI{3}{\cm}, both square with an active area of \SI{100}{\cm\squared}, filled with Ar/CO$_2$(70/30) at atmospheric pressure. Twenty thousand X-ray photons with an energy of \SI{3}{\kilo\electronvolt} (the energy of argon's K$_\alpha$ spectral line) were generated isotropically for each position in a grid with X-Y spacing of \SI{2.5}{\milli\meter}, starting \SI{0.25}{\centi\meter} away from the simulated area's boundary. This was repeated along the Z-axis every \SI{1}{\milli\meter}. The escape probability at each position is determined by the fraction of X-rays leaving the detector volume with respect to the number of generated X-rays. To obtain the exact value, however, one needs to consider the argon fluorescence yield, that is, the probability that a vacancy in the K-shell is filled through a radiative process. The value used in this work was 0.11, obtained using Eq.~1 of~\cite{Kah11}.

\begin{figure}[h]
	\centering
	\includegraphics[width=8cm]{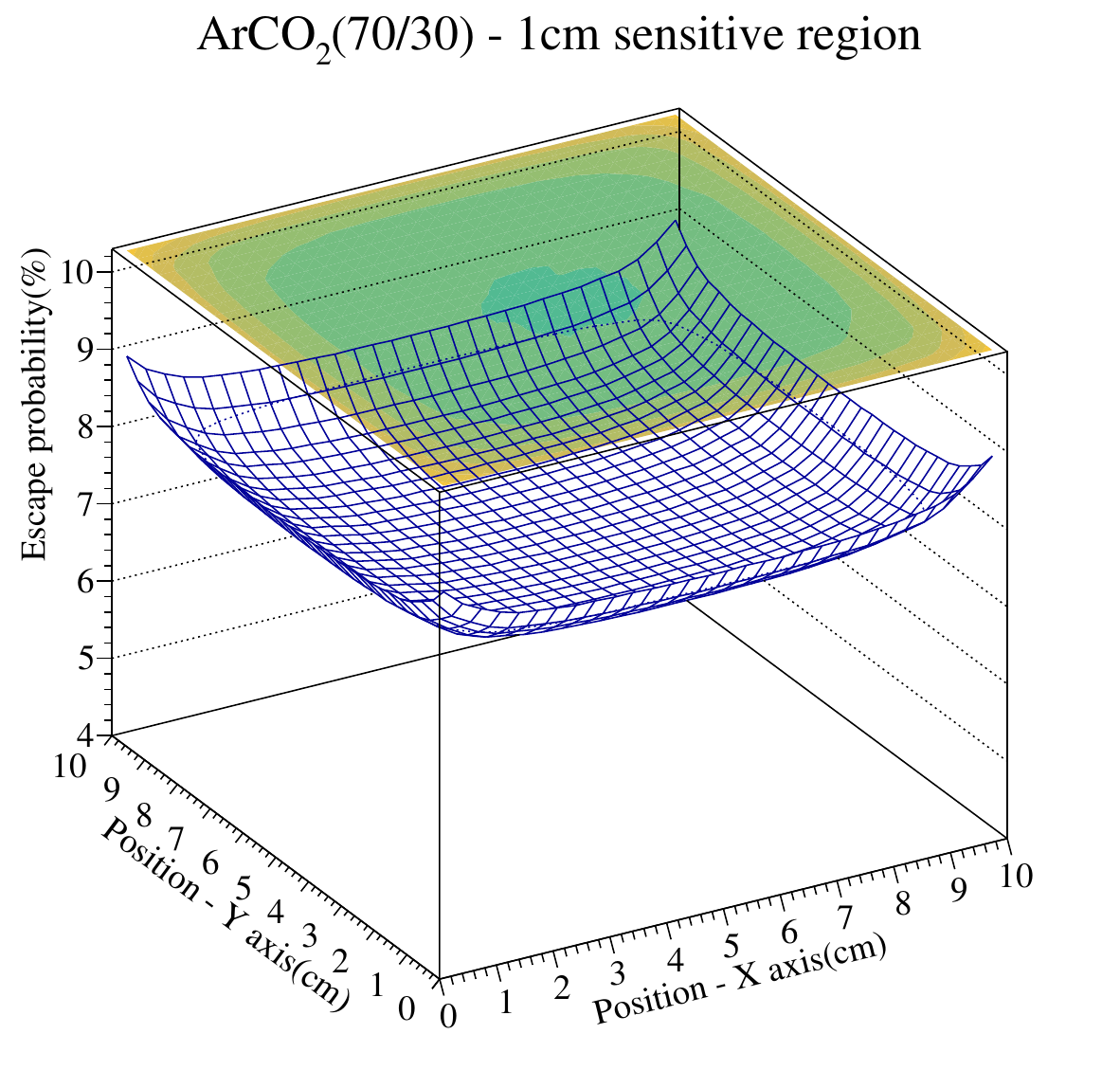} \includegraphics[width=8cm]{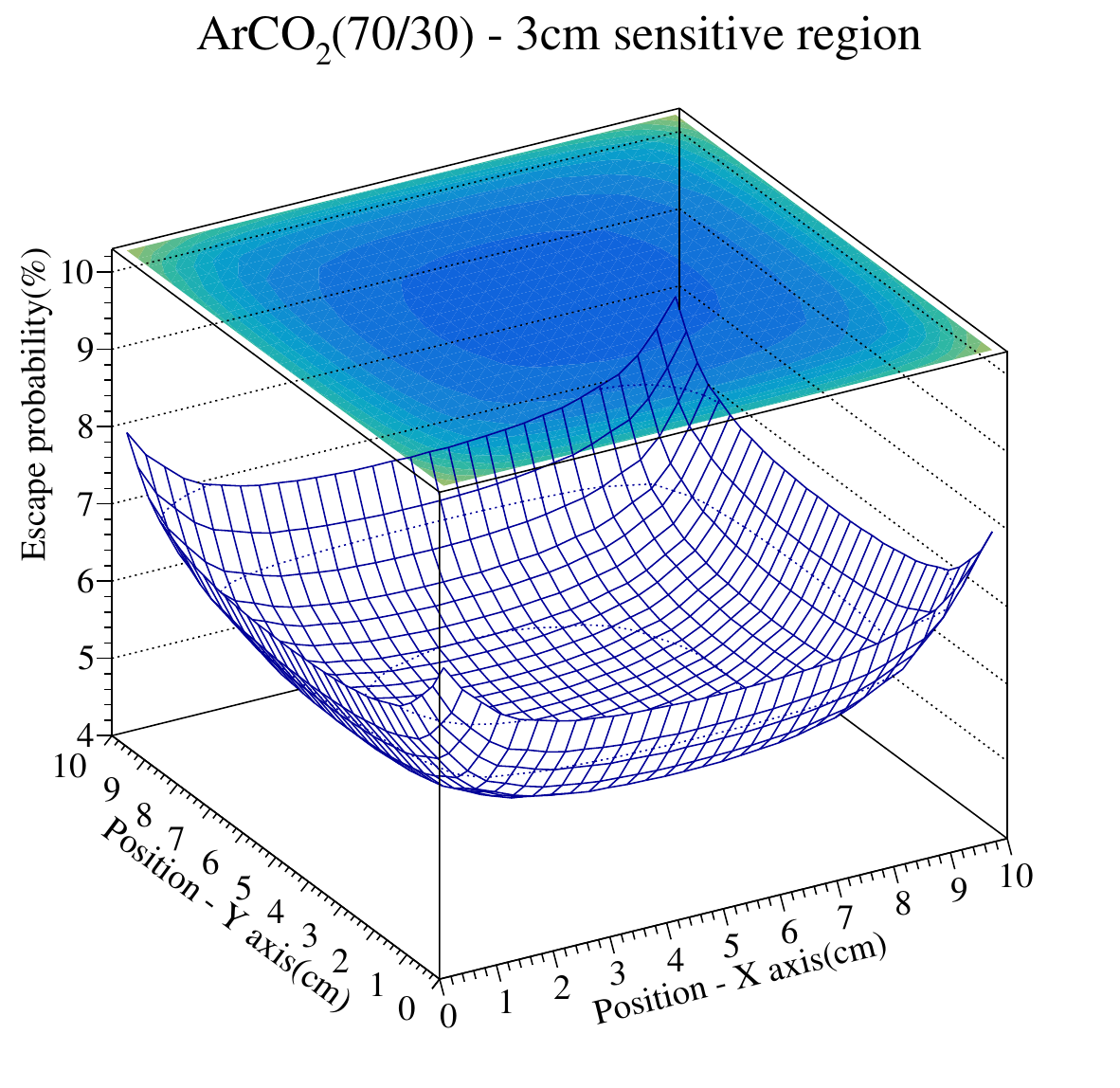} 
	\caption{Escape probability for two different geometries. For the sensitive region with \SI{3}{\centi\meter} the escape probability is lower. As expected, near the edges, the probability of X-rays escaping the detector is higher.}
	\label{escape}
\end{figure}

\begin{figure}[h]
	\centering
	\includegraphics[width=8cm]{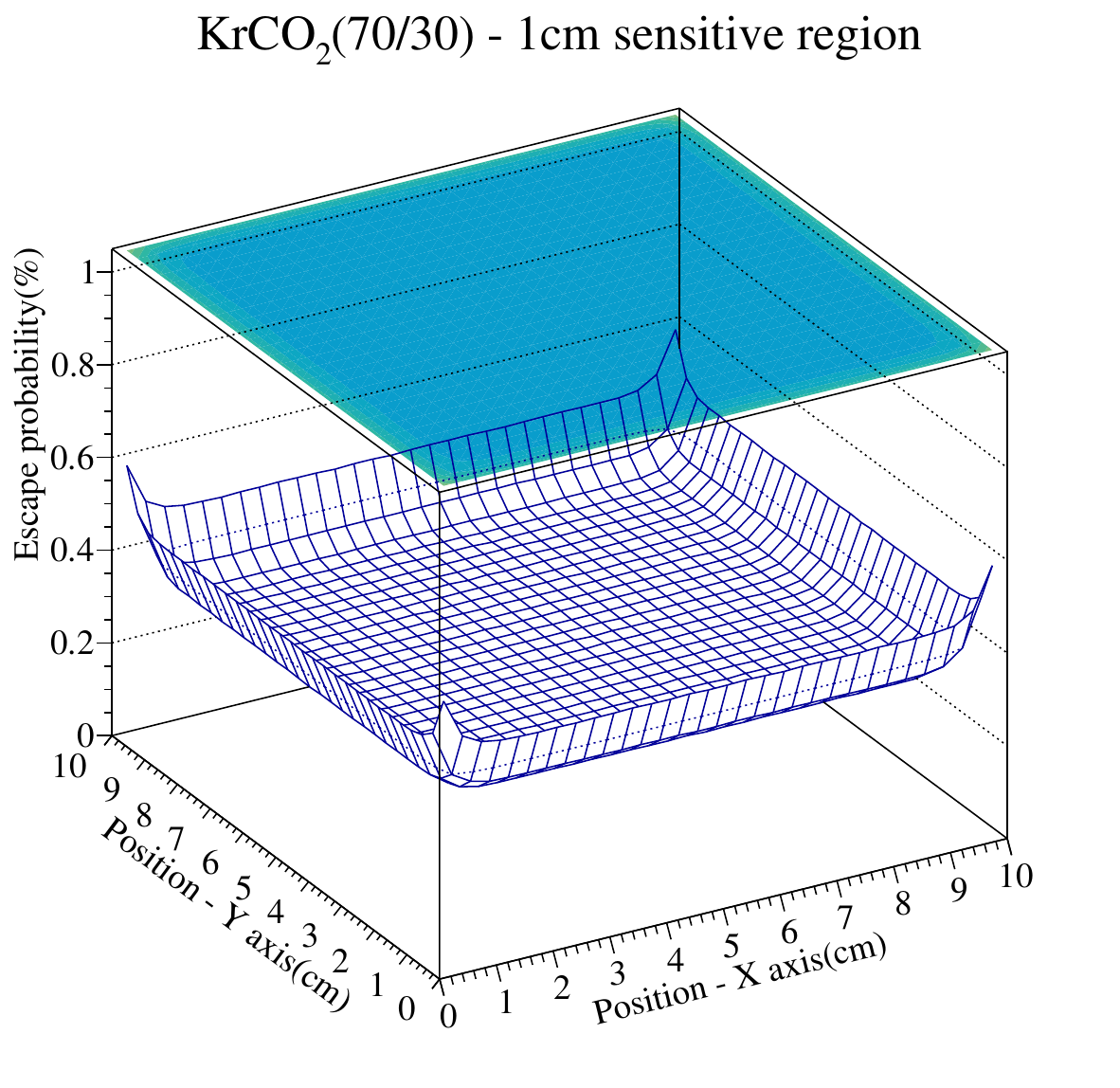}
	\caption{Escape probability for a \SI{1}{\centi\meter} sensitive region gap filled with a krypton based mixture. The escape probability is around 0.4\%, more than one order of magnitude lower than the one with argon.}
	\label{escape_Kr}
\end{figure}

Figure \ref{escape} shows the escape probability for a photon of \SI{3}{\kilo\electronvolt}, argon's K$_\alpha$ fluorescence energy, for different positions in the detector. As the plots show, the escape probability drops from around 7.6\% to around 5.5\% in the central region of the detector only by increasing the depth of the sensitive region by \SI{2}{\centi\meter}. If there is a necessity to work with thinner detectors (smaller depth of the sensitive region), working with gasses with a larger atomic number should be considered, increasing the X-ray absorption cross section and, depending on the incident energies, also decreasing the energy of the secondary photons. Krypton only emits the K$_\alpha$ line if it is ionized by radiation with energy higher than \SI{14}{\kilo\electronvolt}. For lower energies, the L$_\alpha$ transition emits a photon with \SI{1.59}{\kilo\electronvolt}\cite{EmissionLines}. Since this energy is lower, the probability of these photons being absorbed is higher, and the complete energy of the event is deposited in the gas in most of the interactions. Figure~\ref{escape_Kr} confirms that the probability of a secondary photon escaping the detector is much lower than for argon. Similarly to what happened with the argon-based mixture, the escape probability also increases slightly near the edges of the detector. However, because of the much smaller range of the X-rays at such low energy (the radiation length of \SI{1.59}{\kilo\electronvolt} in krypton at 1 bar is approximately \SI{0.28}{cm}), they happen much closer to the boundaries of the detector.

\section{Copper background in the energy spectra}
\label{sec:coppercontamination}
\begin{figure}
	\centering
	\includegraphics[width=10cm]{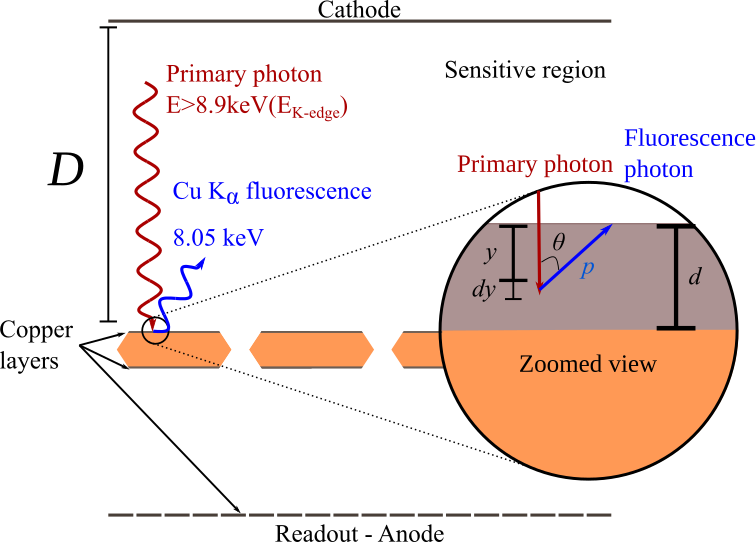}
	\caption{Schematics of the paths traveled by the primaries and fluorescence photons. The copper layer is represented in gray and the kapton in orange. The circular inset is a zoomed view of the small region marked with small circle. $D$ is the depth of the sensitive region of the detector, $d$ is total thickness of the copper layer, $y$ is the distance travelled by the incoming X-ray photon before it is absorbed in the infinitesimally small segment $\mathrm{d}y$, $p$ is the distance travelled by the resulting fluorescence photon after being emitted with an angle $\theta$ with respect to the normal to the surface of the copper layer.}
	\label{contamination}
\end{figure}

One of the challenges regarding the detection of fluorescence X-ray photons with an imaging system based on GEMs is the background added to the energy spectrum generated by the copper electrodes for X-ray energies above \SI{8}{\kilo\electronvolt}. This problem is reported in different works using GEM-based detectors for imaging~\cite{Min17, Min20, Dab16,  Mar19, Car23}. A part of the radiation which does not interact in the sensitive region may hit the copper cladding of the GEM foil or the readout electrode, generating fluorescence X-rays that can enter the absorption region. The characteristic K-shell of copper has an energy of \SI{8.05}{\kilo\electronvolt}, inconveniently close to the efficiency peak of our system, as shown before in figure \ref{gas_eff}. As calculated previously, the energy resolution limit for our detector is around 15\% at this energy range and the presence of this peak may jeopardize the identification of elements. Figure~\ref{contamination} shows how the process happens.

To estimate the contribution of the copper background to the energy spectrum, we adapted a model that is known in XRF spectroscopy \cite{Jak77,Tro92} and implementing new terms which are related with the geometry of our detector. The derivation of the formulas is done taking into account the amount of fluorescence photons created in each electrode of the detector, that is, the top and bottom electrodes of each GEM foil and the readout plane. The final result is the sum of all these contributions. 

The fluorescence rate $\mathrm{d}I_k$ that comes from the fluorescence K$_\alpha$ line of copper and is emitted by the infinitesimally thin layer d$y$ of the top copper cladding of the first GEM foil can be obtained as following:

\begin{equation}
	\mathrm{d}I_K(E_i, E_K) \propto I_0(E_i)\frac{\mathrm{d}\Omega}{4\piup} \omega_K \frac{1}{\lambda_K(E_i)} \exp\left(-\-\frac{y}{\lambda_\mathrm{Cu}(E_i)}\right)\exp\left(-\frac{p}{\lambda_\mathrm{Cu}(E_K)}\right)\mathrm{d}y 
    \label{eq:dI}
\end{equation}

\noindent where $I_0(E_i)$ is intensity of the incident photon with energy $E_i$, $\Omega$ is the solid angle, $y$ is the depth in the copper layer where the interaction occurred, $p$ is the distance crossed by the fluorescence X-ray before leaving the copper layer. As before, this equation results from the direct use of the Beer-Lambert's law that determines the probability of a photon being absorbed at a given depth in a material. The $y$ and $p$ variables are depicted in figure \ref{contamination}. $\omega_K$ is the fluorescence yield of copper, i.e., the probability of the K-shell vacancy being filled through radiative processes. The value used was 0.433, which was calculated using Eq.~1 of~\cite{Kah11}. The term $\lambda_K(E_i)^{-1}$ is proportional to the fraction of interactions that generate K-shell vacancies. In a first approximation, at these energies, the difference between the attenuation length using all the electronic shells or only the K-shell is small and can be neglected~\cite{Hir00}, leading to $\lambda_K(E_i)^{-1} \approx \lambda_\mathrm{Cu}(E_i)^{-1}$. The first exponential term gives the intensity of the beam (with energy $E_i$) that reaches the infinitesimal layer of copper d$y$, after traveling a copper length $y$. The second exponential is the attenuation of the fluorescence photons (with energy $E_K$) that are generated inside this same infinitesimal layer that travel the distance $p$ inside the copper layer. 

The probability of fluorescence photons being detected $\mathrm{d}\bar{I}_K$ is obtained by multiplying the probability of fluorescence $\mathrm{d}I_K$ by the probability of transmission of the incoming photons through the sensitive region without interacting with the air outside the detector, the window, the cathode material or the gas $\gamma(E_i)$; and the probability of interaction of the fluorescence photon, emitted at an angle $\theta$ with respect to he normal to the surface of the GEM foil, within the depth of the sensitive region:
	\begin{equation}
	\mathrm{d}\bar{I}_K(E_i, E_K) = 
    \mathrm{d}I_K(E_i, E_K) \cdot 
 \gamma(E_i) \cdot \left(1-\exp\left[\frac{-D}{\lambda_\mathrm{Ar}(E_K)\cos(\theta)}\right]\right),
	\label{eq:conta1a}
	\end{equation}
where the term in round brackets is simply the probability of absorption in the gas, $D$ is the depth of the sensitive region and $\lambda_{Ar}(E_K)$ is the attenuation length of X-rays in argon.

Combining Eqs.~\ref{eq:dI} and \ref{eq:conta1a}, Eq.~\ref{eq:conta1} is obtained, where $p$ is rewritten as $p = \frac{y}{\cos(\theta)}$:

\begin{small}
	\begin{equation}
	\mathrm{d}\bar{I}_K(E_i, E_K) = \frac{ I_0(E_i)\mathrm{d}\Omega\gamma(E_i) \omega_K}{4\pi\lambda_{Cu}(E_i)} \exp\left[-\left(\-\frac{y}{\lambda_{Cu}(E_i)}+\frac{y}{\lambda_{Cu}(E_K)\cos(\theta)}\right)\right]\left(1-\exp\left[\frac{-D}{\lambda_{Ar}(E_K)\cos(\theta)}\right]\right)\mathrm{d}y 
	\label{eq:conta1}
	\end{equation}
\end{small}

The total fluorescence intensity from the first copper layer of the first GEM foil can be obtained by integrating over the thickness of the copper cladding $d$ for the whole solid angle (with $\theta$ and $\phi$ denoting the polar and azimutal angles, respectively):

\begin{align}
	\begin{split}
		\bar{I}_K(E_i, E_K) &= \frac{ I_0(E_i)\gamma(E_i) \omega_K}{4\pi\lambda_\mathrm{Cu}(E_i)}\int_{0}^{2\piup}\mathrm{d}\phi \int_{0}^{\frac{\piup}{2}}\mathrm{d}\theta \sin(\theta)\int_{0}^{d}\mathrm{d}y  \exp\left[-\left(\-\frac{1}{\lambda_\mathrm{Cu}(E_i)}+\frac{1}{\lambda_\mathrm{Cu}(E_K)\cos(\theta)}\right)y\right] \\
		&\qquad 
		\left(1-\exp\left[\frac{-D}{\lambda_\mathrm{Ar}(E_K)\cos(\theta)}\right]\right),
	\end{split}
\end{align} 
leading to

\begin{small}
	\begin{equation}
	\bar{I}_K(E_i, E_K)=\frac{ I_0(E_i) \gamma(E_i) \omega_K }{2}    \int_{0}^{\frac{\piup}{2}}\mathrm{d}\theta\sin(\theta)
	\frac{a}{a+b\cos(\theta)}\left(1-\exp\left[-\left(a+\frac{b}{\cos(\theta)}\right)d\right]\right)\left(1-\exp\left[\left(\frac{-Dc}{\cos(\theta)}\right)\right]\right),
	\label{kappa_detection}
	\end{equation}
\end{small}
where , $\lambda_{Cu}(E_i)^{-1}$, $\lambda_{Cu}(E_K)^{-1}$ and $\lambda_{Ar}(E_f)^{-1}$ are denoted as $a$, $b$ and $c$, respectively. This general derivation is taking only the first layer of copper into account. The same reasoning can be applied for the second copper layer of the foil or the layers in subsequent GEM foils and the readout electrode, taking into account the probabilities of transmission through all the previous layers.

This integral can be numerically solved assuming some reasonable approximations. The first one is that the X-ray source is considered to be far from the detector window, meaning that all the photons are entering the sensitive region perpendicularly to the detector plane. The detector is considered to have an infinite area, and all the layers on the GEM foils and readout are considered as continuous electrode planes. The absence of copper in the holes of the foils is neglected as well as the spaces between the strips of the readout plane. It can be shown that the lower fluorescence probability caused by the smaller amount of copper in the first layer is actually compensated by the less shielding against the fluorescence radiation originating from the other copper layers. These assumptions are approximations that simplify the problem with small changes in the final result. The values for the attenuation length used were obtained using the tabulated values \cite{XCOM, Hen93}.

\begin{figure}[h]
	\centering
	\includegraphics[width=10cm]{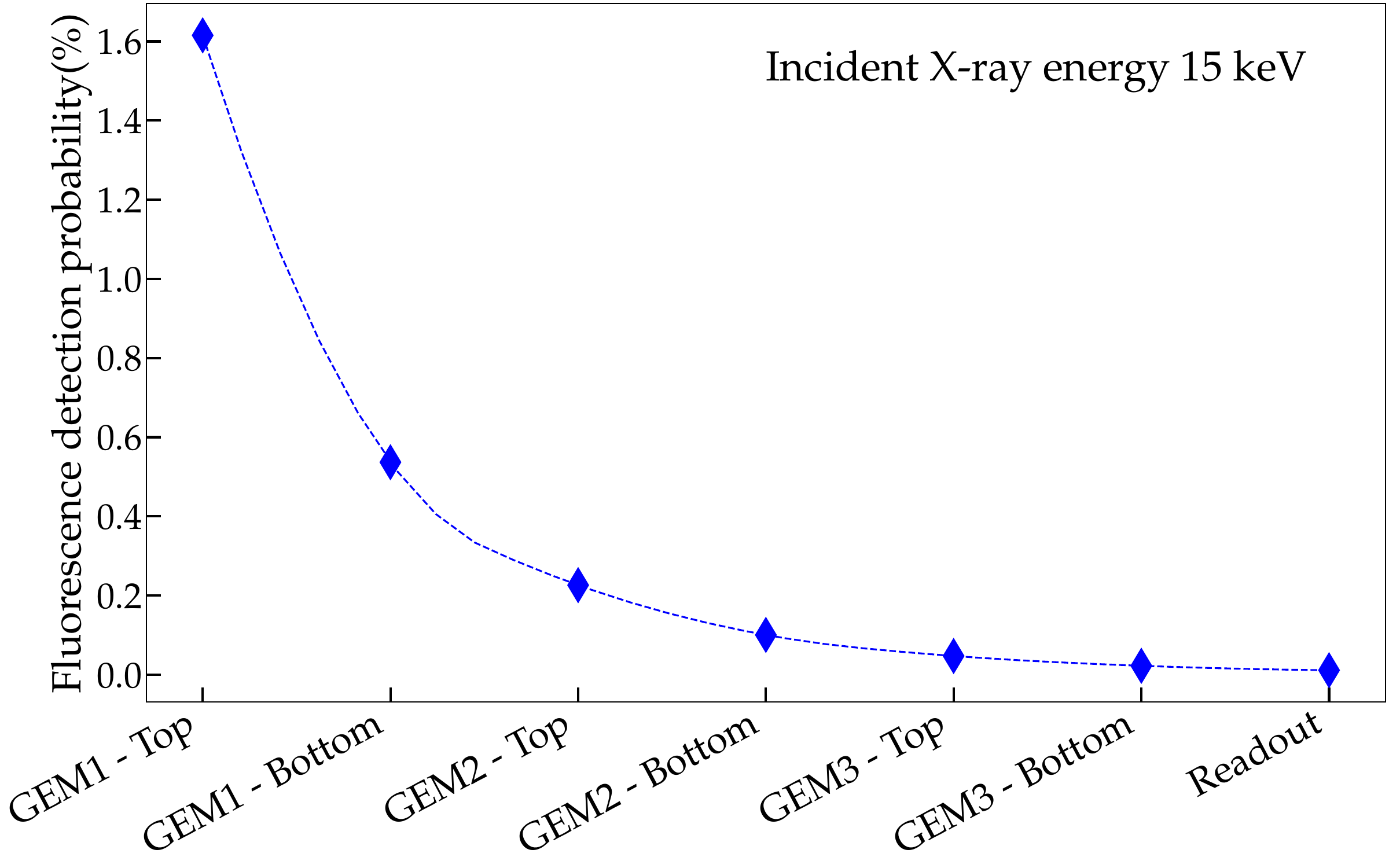}
	\caption{Contribution to fluorescence generated by each copper layer of the detector.}
	\label{contamination_contribution}
\end{figure}

Figure \ref{contamination_contribution} shows the probability of detection of the fluorescence X-rays emerging from the different copper layers inside a detector. The probabilities, were calculated using Eq.~\ref{eq:conta1} and applying the additional terms for each copper layer as described. As expected, the contamination of energy spectra with fluorescence photons that were generated in layers which are deeper inside the detector are smaller. 

\begin{figure}[h]
	\centering
	\includegraphics[width=8cm]{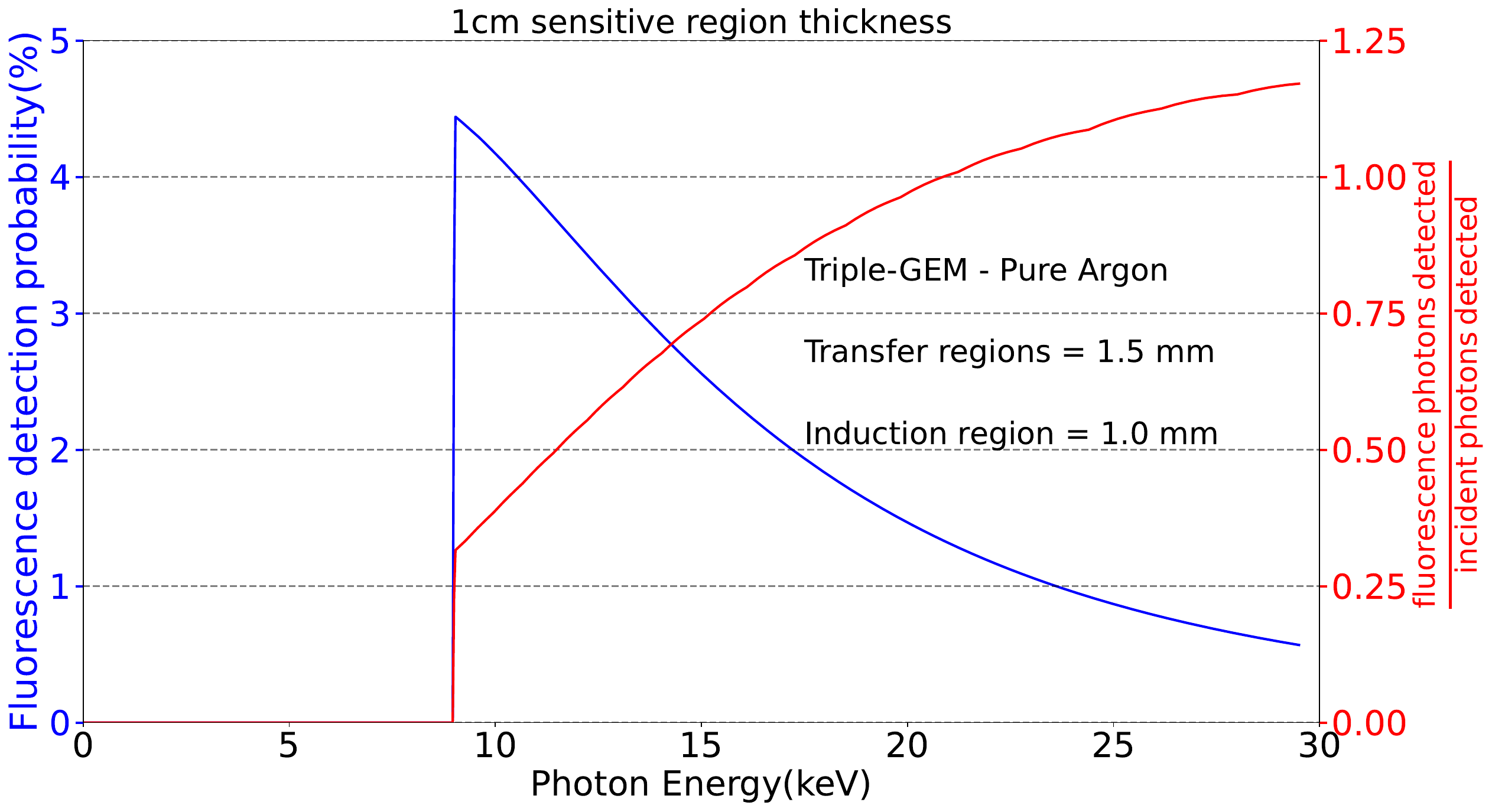} \includegraphics[width=8cm]{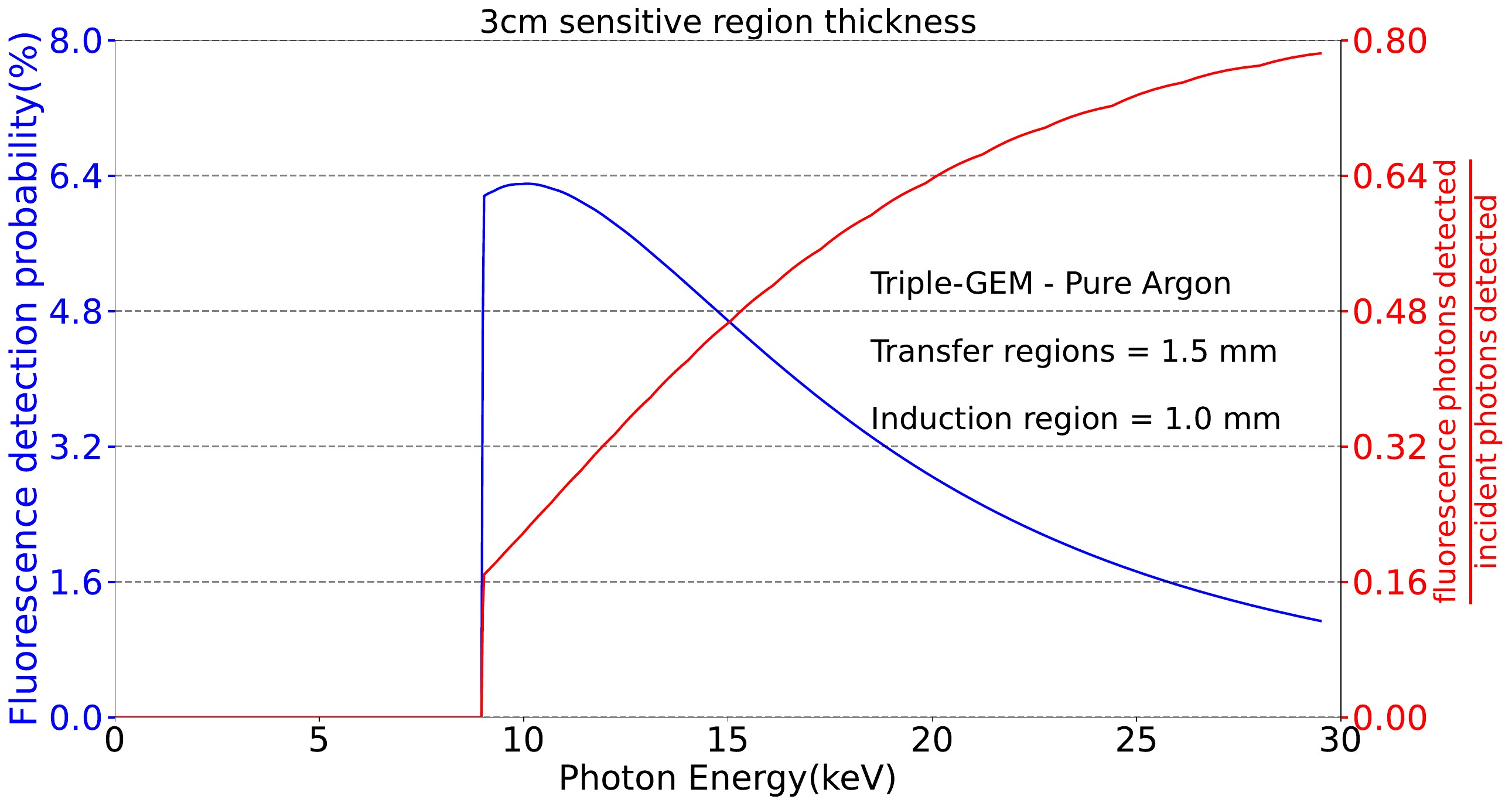}
	\caption{Copper fluorescence detection probability (blue) 
		and fluorescence detection fraction (red) for 1 cm and 3 cm sensitive region thickness.}
	\label{gem_contamination}
\end{figure}

Figure \ref{gem_contamination} shows the total copper fluorescence detected for two different systems. The one on the left side has a sensitive region with \SI{1}{\centi\meter}, while the second one, on the right side, has a sensitive gap with \SI{3}{\centi\meter}. Both detectors have three GEM foils and a copper coated readout. Each copper layer has a thickness of \SI{5}{\micro\meter}. The transfer region~(gap between GEM foils) is \SI{1.5}{\milli\meter} deep and the induction region~(gap between the last GEM foil and the readout plane) is \SI{1}{\milli\meter}.  The value used for $\gamma$ in Eq.~\ref{kappa_detection} to compute the first layer contribution was kept equal to the transmission probability in the sensitive region. For the second and third copper layers of the detector, the attenuation of the incident X-rays by the components that are above them were taken into consideration (for example, a intensity of photons generated in the top layer of the second GEM foil can be attenuated by the transfer region and the first GEM foil of the stack).

One can see that the probability of detecting fluorescence from copper (blue curve) has its maximum just above the K-shell absorption edge and then it drops. This happens mainly due to the increase of the attenuation length for higher energy primary photons, while the attenuation length for the copper K$_\alpha$ photons remains constant. In red (secondary Y-axis) it is possible to see the fraction of fluorescence photons detected with respect to the number of the original incident photons. Increasing the number of incident photons which are detected reduces the effect of the contamination of the fluorescence spectra with the copper emitted by the detector. Therefore, it is desirable to have the ratio curve scanning lower values. Hence, the detector with thicker sensitive region shows more advantages. It is simple to understand that this contamination of the spectra also worsens the position resolution of the detector on one hand (these photons create a flat background in the images, decreasing the contrast over the whole active area) and, on the other hand, generate an excess of signals at \SI{8.05}{\kilo\electronvolt} plus the correspondent argon escape peak at \SI{5}{\kilo\electronvolt}. 

\begin{figure}[h]
	\centering
	 \includegraphics[width=8cm]{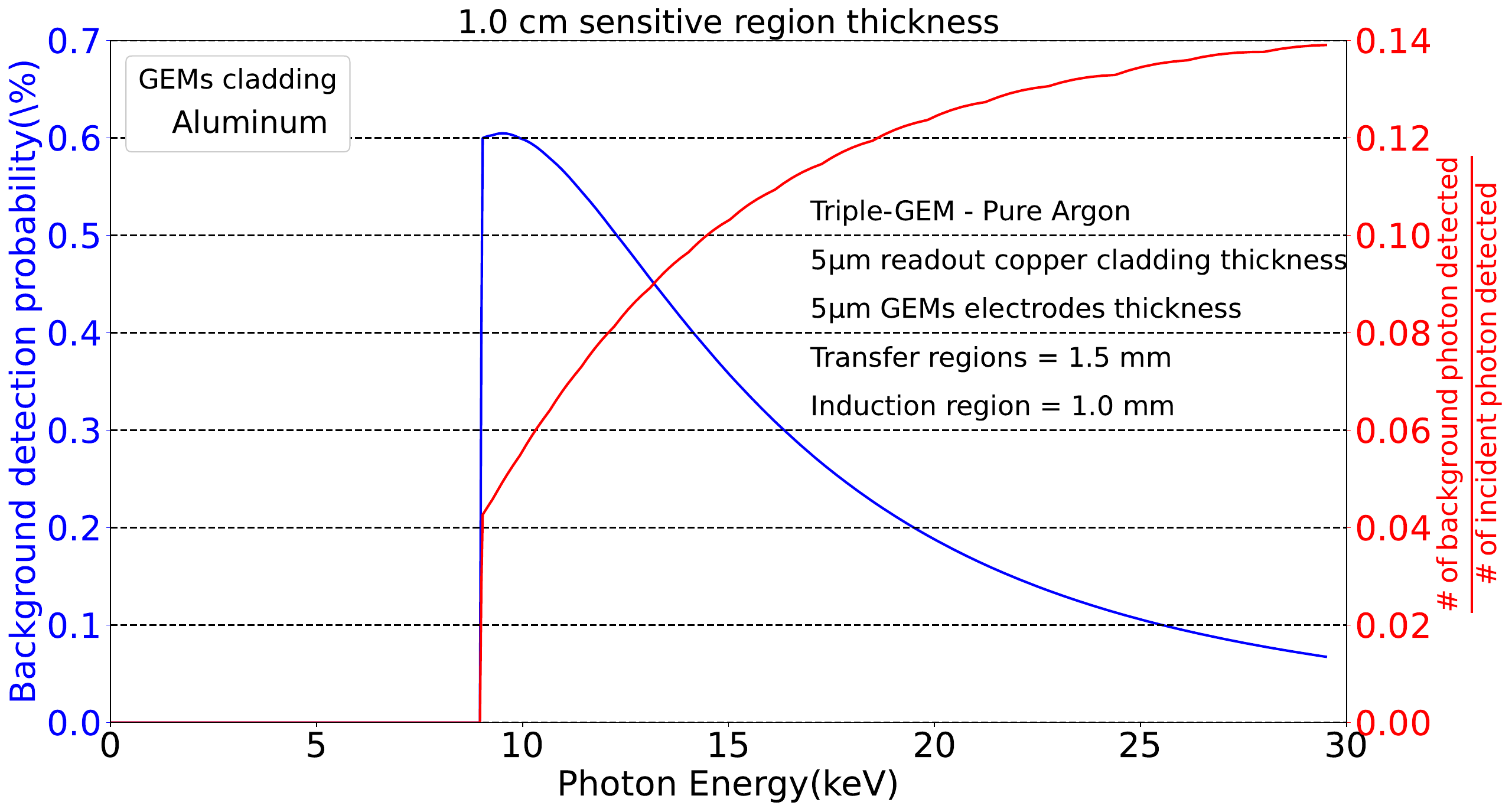}
	\caption{Copper fluorescence detection probability for a triple-GEM detector using aluminum cladding and copper readout.}
	\label{readout_contamination}
\end{figure}

To cope with this issue, different groups are working in methods to suppress the fluorescence of copper in GEM-based detectors. One idea is to remove most of the copper conductive layer and to use only the adhesive which glues the copper to the dielectric as a conductor \cite{Min17, Min20}. This layer has chromium in its composition, which can also produce a fluorescence, however, it is much thinner and should not affect the energy spectrum as much as the standard conductive layer made of copper. Another alternative is to use a different cladding for the GEM foils, replacing the copper by another metal with a smaller atomic number. Aluminum, with X-ray fluorescence at \,\SI{1.48}{\kilo\electronvolt} which will not affect the energy spectrum in the region of interest, is also under study by other groups \cite{Mar19, Car23}. Our calculations also follow the trend of those experimental results, as seen in figure \ref{readout_contamination} that shows the total fluorescence calculated for a aluminum-clad triple-GEM detector operating with a copper readout, with the ratio curve at much lower values, indicating a much larger amount of incident photons with respect to the photons from fluorescence in the detector.

The relative dimensions were kept the same as the ones for the previous configurations and there is no possibility of fluorescence of aluminum, because only photons with energies well above the Al fluorescence threshold are able to reach the sensitive region. One can see that the probability of generating a fluorescence photon of copper is small in the system using aluminum GEMs because only the copper present in the readout is contributing for this background. Although this is a promising result, the etching of aluminum foils is more difficult and is still in development. Furthermore, these electrode seem to be more prone to spontaneous discharge problems in a shorter time than the standard copper-clad foils~\cite{Mar19}.

\section{Conclusion and outlook}

In this work, a few common issues found in GEM-based detectors when operating in X-ray imaging and energy dispersive applications were studied and described with the help of simulations and analytical calculations that use basic concepts of radiation Physics. A comparison of the results obtained with existing experimental work was done whenever possible. Even in the absence of systematic published experimental work (for example in the influence of the gas mixture and size of the sensitive region in the escape peaks), common understanding of gaseous detectors allows to understand the results that were obtained. The aim of this work is the provision of simple models that deepen the understanding in GEM-based detectors for imaging applications, validating them with the extended knowledge that was acquired since their invention in 1997. The tools presented in this work are extendable to gaseous detectors in general, provided that the equations and simulations are properly adapted to the geometry. In fact, concepts such as the distributions of the number of primary electrons, multiplication factors, escape peaks and fluorescence of the detector materials are well know long before the existence of MPGD. 

GEM detectors have shown their importance in high energy physics experiments and they also may be used for other applications such as XRF, PIXE and X-ray diffraction if the correct parameters, geometry and materials are chosen. The main advantage of using such detectors is connected to the large sensitive areas that are possible to manufacture, without prohibitively increasing their price and that can dramatically reduce the scanning time of large samples. For applications such as XRF it was shown that the spectral analysis may present some challenges due to the limitations in energy resolution which does not allow distinguishing neighboring elements in the periodic table.

The detection efficiency can achieve larger values and the escape peak interference may be suppressed at a point that it stops influencing the final spectrum by increasing the depth of the sensitive region or by changing the gas according to the specific application. Everything will depend on the needs of the experiment and availability of material.

Regarding the copper fluorescence background in the energy spectrum, it was possible to evaluate the probability to detect undesired fluorescence photons and estimate the the detector background. Showing that detectors with larger sensitive regions or without copper in its composition will be more robust against the contamination of the energy spectra with fluorescence X-rays from the copper surface of the GEM foils. 
 
\newpage
\section*{Acknowledgments}
 
This study was financed in part by the Coordenação de Aperfeiçoamento de Pessoal de Nível Superior – Brasil (CAPES) – Finance Code 001, by the grant 2016/05282-2 from Fundação de Amparo à Pesquisa do Estado de São Paulo, Brasil. H. Natal da Luz acknowledges support from GAČR grant GA21-21801S (Czech Science Foundation).

\bibliography{mybibfile}

\end{document}